\documentclass{article}

     \usepackage[final,nonatbib]{neurips_2020}

\usepackage[utf8]{inputenc} 
\usepackage[T1]{fontenc} 
\usepackage{hyperref} 
\usepackage{url} 
\usepackage{booktabs} 
\usepackage{amsfonts} 
\usepackage{nicefrac} 
\usepackage{microtype} 
\usepackage{graphicx} 
\usepackage{subcaption}
\usepackage{siunitx}
\DeclareUnicodeCharacter{2009}{\,}

\usepackage{amsmath, bm}
\usepackage{amssymb}
\usepackage{color}
\usepackage{xcolor}
\definecolor{redcol}{rgb}{1, 0, 0}
\definecolor{bluecol}{rgb}{0, 0, 1}

\title{Transformer based Molecule Encoding for Property Prediction}

\author{
    Prateeth Nayak 
    \And
    Andrew Silberfarb 
    \And
    Ran Chen
    \AND
    Tulay Muezzinoglu
    \And
    John Byrnes 
    \AND

    {\normalfont SRI International}\\ San Diego, CA \\ \\
    \texttt{\{first.last\}@sri.com}

 }

\begin{document}

\maketitle

\begin{abstract}
  Neural methods of molecule property prediction require efficient encoding of
  structure and property relationship to be accurate.  Recent work using graph
  algorithms shows limited generalization in the latent molecule encoding
  space. We build a Transformer-based molecule encoder and property predictor
  network with novel input featurization that performs significantly better
  than existing methods. We adapt our model to semi-supervised learning
  to further perform well on the limited experimental data usually available in practice.

\end{abstract}

\section{Introduction}
Much of drug discovery focuses on finding a molecule that binds to a certain
protein site, does not bind to other sites, and has chemical properties that
allow it to be distributed into the cells by the body.  Synthesizing and
analyzing compounds is extremely expensive, so the ability to predict these
properties using machine learning has obvious value.  In recent years, a number
of representations of molecules have been suggested which are suitable for
machine learning of molecular properties, the most successful of which are
based on message passing (MPNN's).  Unfortunately, these architectures require
large amounts of training data, but for many chemistry problems only a small
number of molecules with the desired properties may be known.  Self-attention
provides a richer and more efficient neural architecture than an MPNN, allowing
the network to learn which atoms to pass messages between. To enable this we
adapt the standard transformer architecture to better suit molecular
processing, allowing it to make use of bond information and information about
uncertainty.  We add to this prediction of computable properties, which allows
us to augment a set of molecules having assay data with arbitrary additional
molecules for which only the computable properties are known.  The result is a
semi-supervised model that learns an improved encoding of structure, achieving
15-20\% error reduction over MPNN's on small data sets and close to 10\% error
reduction over the best known algorithms on full data sets.

\section{Related Research}
\label{related}

Most work on property prediction in recent years has followed one of two paths:
(1) use of better and well-formulated molecule descriptors/fingerprints such as
ECFP\cite{ecfp};
(2) use of novel model architectures that operate on SMILES strings or
structural graphs.
%
%
We propose a new approach in the latter path that combines the
state-of-the-art Transformer technique originally used for language translation
\cite{transformer,transformerXL,shaw18}
with message passing within molecular graphs \cite{moleculenet}, which have
become the benchmark for state-of-the-art performance.
\cite{Tang} and \cite{mit} have already shown improvement over
\cite{moleculenet} through application of self-attention to message-passing.
We extend this work by modifying the attention mechanism in existing
Transformer networks to help encode a fixed sequence for representation of
molecule, improving the performance further.
\cite{moleculenet},\cite{mit} evaluate their work on extensive public and private
datasets; we focus entirely on the datasets Lipophilicity \cite{moleculenet} and
ESOL\cite{esol}.
\cite{smilessyn} and \cite{shin} apply Transformer networks to SMILES, a
sequential symbolic encoding of the molecule graph, but this work attempts to
capture the symbol sequence and not the implied graphical structure that the
sequence represents.  To the best of our knowledge, ours is the first work that
highlights the importance of performance on datasets that contain relatively
few labeled samples, but where large numbers of candidate molecule features are
available.

\section{Approach}
\label{approach}

We use a modified Transformer\cite{transformer} neural network to predict molecule properties. We treat a molecule as a set of atoms, each atom is represented by a vector that depends on its chemical properties (e.g. valence, periodic table row, column, etc.). We do not use a positional encoding, such that permutations of the input atoms within a molecule are irrelevant.  For relational features we encode bond features between pairs of atoms, and use this bond information to modify the Transformer attention mechanism by building on the relative position approach used in TransformerXL \cite{transformerXL}.  This modification emulates the action of a Message Passing Neural Network in Gilmer et al.\cite{MPNN}, using the self-attention mechanism of the Transformer and is similar in spirit to Tang et al.\cite{Tang}.   We also include a standard-deviation term in the value update of the Transformer, allowing the model to make use of information about the uncertainty of the feature vector over the attended atoms at each layer. Notionally, this allows subsequent layers of the transformer to pay more (or less) attention to regions of the molecule with high variance, or extreme values.

\paragraph{Model Input Featurization}
Our featurization of a molecule starts with the Weave features \cite{weave}. Each atom in the molecule is associated with a set of categorical features, each value of which is associated with a learnable vector of dimension $256$. The vectors for all of the features are added together to form the input vector to the Transformer. During training the number of atoms is padded to a fixed length $50$ and the standard Transformer masking mechanism is used to assign $0$ self-attention weights to these missing atoms at each layer. Each pair of atoms is also associated with a bond type given by the same Weave featurization. The bond type for each atom pair is provided as an additional input to the Transformer self-attention mechanism as described below. We modify the Weave featurization; we break atom element up into two features, periodic table row, and periodic table column. We add a set of binary features for each atom derived from RDKit's low level pattern matching for Aromaticity, Hydrogen Bond Acceptor and Donor, Zinc Binding Atom, Hydrophobe and Lumped Hydrophobe. These are computed using the family designation from an RDKit FeatureFactory based on 'BaseFeatures.def'. Finally we add an additional feature for each atom, corresponding to the number of obscuring atoms. Obscuration is defined according to the RDKit force directed layout, and is the count (0, 1, 2 or more) of other atoms that lie opposite the direction of highest local density.  Mathematically an atom with position $\vec{y}$ obscures an atom with position $\vec{x}$ if $(\vec{y} - \vec{x}) \dot{} \vec{r} < 0$ with $\vec{r} = \sum_z (\vec{z} - \vec{x}) \exp^{-|\vec{x}\cdot\vec{z}|^2 / 8.0}$.

\paragraph{Model Architecture}
Starting with the basic Transformer self-attention mechanism, we add bond based relational attention. Each pair of atoms is labeled with a bond type $b_{ij}$. At a given transformer layer each atom $i$ will have associated query, $q_i$, and key, $k_i$, giving scaled dot-product attention $a_{ij} = \text{softmax}(q_i \cdot k_j / \sqrt{\text{dim}})$. We modify this by associating to each bond value $b$ a weight $w_b$, a query vector $q_b$ and a key vector $k_b$.  These bond weights and vectors are learned independently for each head in each layer of the Transformer, allowing each Transformer head to focus on a different aspect of the bonds as needed. Note that for brevity and clarity all equations are presented for a single transformer head. The full self-attention formula is then
\begin{equation}
  a_{ij} = \text{softmax}_j\left. \left[\alpha (q_i \cdot k_j + q_b \cdot k_j + q_i \cdot k_b) + w_b \right] \right|_{b=b_{ij}},
\end{equation}
where we use $\text{softmax}_j$ to denote application of softmax over dimension $j$.

The standard self-attention based update at layer $l$ is $x_i^{l+1} = x_i^l + \text{agg}(\sum_j a_{ij}^l v_j^l)$. Here $x_i^l$ is the vector for atom $i$ at layer $l$, and $v_j^l$ is the corresponding value at layer $l$ derived in the standard way as a linear layer applied to $x^l$ for each head. The operator $\text{agg}()$ takes all of the outputs of the self attention heads, concatenates them and passes them through a linear layer to produce an output of the same dimension as the input $x_i^l$. Since $a^l_{ij}$ is the output of a softmax operation it is a valid probability distribution allowing one to define the updates for each atom $i$ as an element wise expectation $E^l_i[v] \equiv \sum_j a^l_{ij} v^l_j$. We then define the corresponding standard deviation $\sigma_i^l[v] = (E^l_i[v^2] - E^l_i[v]^2)^{1/2}$, and use this to create a modified update rule $ x_i^{l+1} = x_i^l + \text{agg}(E_i[v], \sigma_i[v])$ where we now aggerate over both the means and the standard deviations. After passing through the Transformer we are left with a set of atom vectors which need to be combined into a prediction about molecular properties. We first filter out vectors corresponding to empty atoms. Following the approach used in MPNN's we aggregate using symmetric operations: the mean, min, max and standard deviation of the atoms vectors. In addition to these symmetric statistics we also make use of a "pseudo-atom" in the transformer, similar in function to BERT's initial token\cite{bert}, providing the transformer the ability to aggregate and utilize information about the whole molecule, rather than just its atoms. Here the input representation of the pseudo atom is freely learned, while each atom shares two directed pseudo-bonds with the pseudo atom. The output vector for the pseduo atom is treated as if it were a fifth symmetric statistic. We concatenate the output of all five statistics and pass them through a linear layer to create a single vector representing the full molecule. This is then passed through an residual MLP, and necked down using a final linear layer to create the output prediction(s).

\begin{figure*}[t!]
  \centering
  \begin{subfigure}[t]{0.5\columnwidth}
    \centering \includegraphics[scale=0.2]{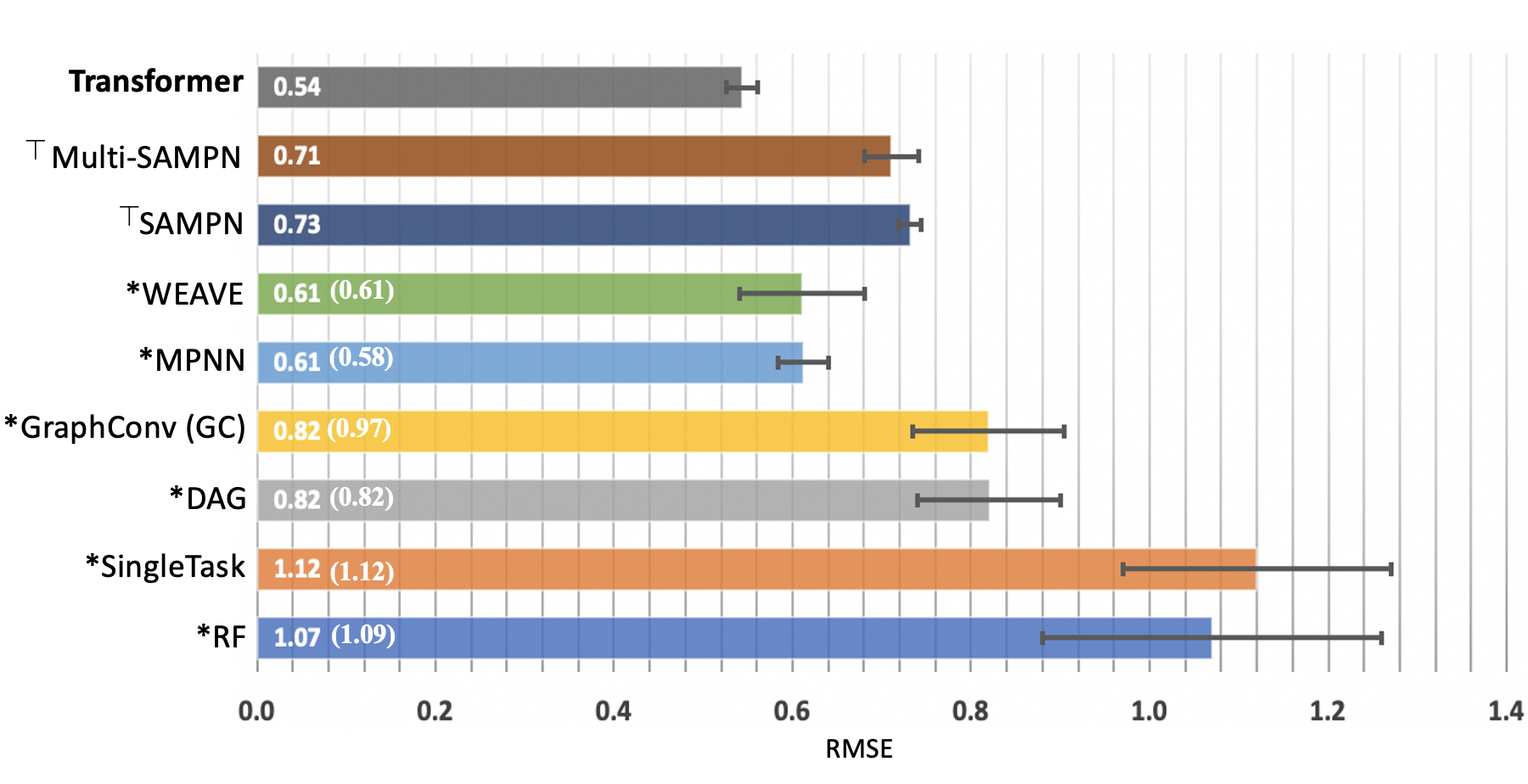}
    \caption{ESOL}
    \label{esol_rmse}
  \end{subfigure}%
  ~
  \begin{subfigure}[t]{0.5\columnwidth}
    \centering \includegraphics[scale=0.2]{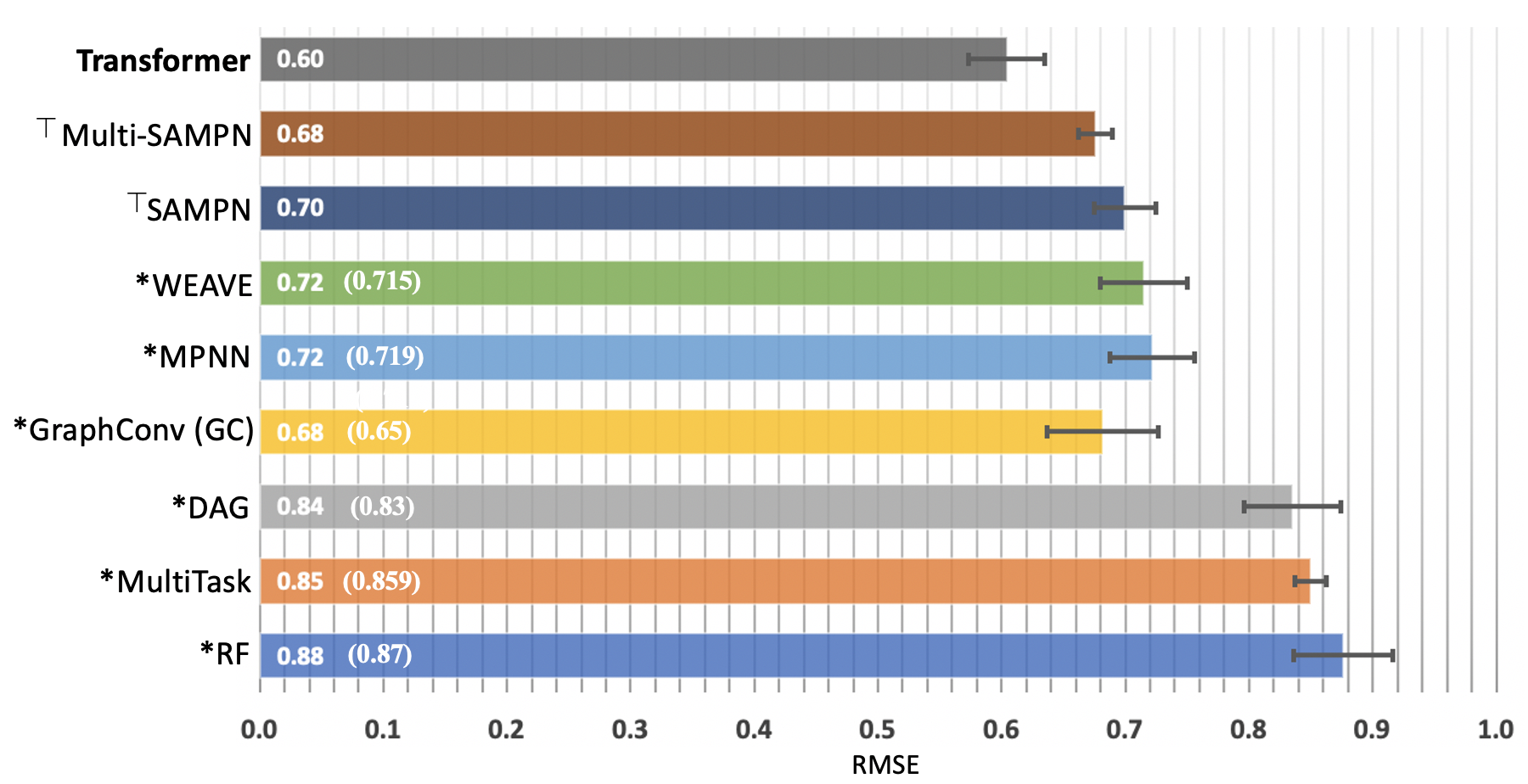}
    \caption{Lipophilicity}
    \label{lipo_rmse}
  \end{subfigure}%
  \caption{Test RMSE results on ESOL and Lipophilicity datasets}
  \footnotesize{\small\textasteriskcentered{ models from Wu et
      al.\cite{moleculenet}}. $\top$ models from Tang et al.\cite{Tang}}
  \label{test_rmse}
\end{figure*}


\paragraph{Semi-Supervised Learning}
In addition to the target property we also predict \textit{Computational Partition Coefficient (cLogP)} and \textit{Molar Refractivity (MR)}, which can be computed for each molecule.  We predict from each atom vector $x_i$ the input features for atom $i$ and the sum of the $x_j$'s for each atom $j$ to which $i$ is bonded by each bond type.  These targets force each atom vector to reflect the local chemistry, providing better generalization for predicting molecule properties.  Since these properties are available for arbitrary molecules, even when laboratory data is not available, we are able to carry out semi-supervised learning where we include unannotated molecules during training.

\section{Experimental Setup and Results}
\label{results}
We evaluate our model on the Lipophilicity and ESOL prediction tasks from \cite{moleculenet}.  Lipophilicity provides
octanol/water partition coefficient (logP) for 4200 molecules and is used in \cite{moleculenet},\cite{mit} and
\cite{Tang}.  ESOL \cite{esol} provides water solubility (logS) for 1128 molecules and is used in \cite{moleculenet}.
We use the scaffold-based data splits specified in \cite{moleculenet}, which are 80\% train, 10\%
validation, and 10\% test.  The validation split is used for hyper-parameter tuning and the test split for reporting.
We repeat each experiment with three different seed values.
We also evaluate performance in low data scenarios by retaining annotations only for subsets of the training data (of
sizes 50, 100, 200, 500, and 900) and reporting RMSE on a fixed test set of size 200.  We use semi-supervised training,
allowing our model to train on those training molecules for which the annotations have been ablated, but not on the
molecules from the test set.
\begin{figure*}[t!]
  \centering
  \begin{subfigure}[t]{0.5\columnwidth}
    \centering \includegraphics[scale=0.4]{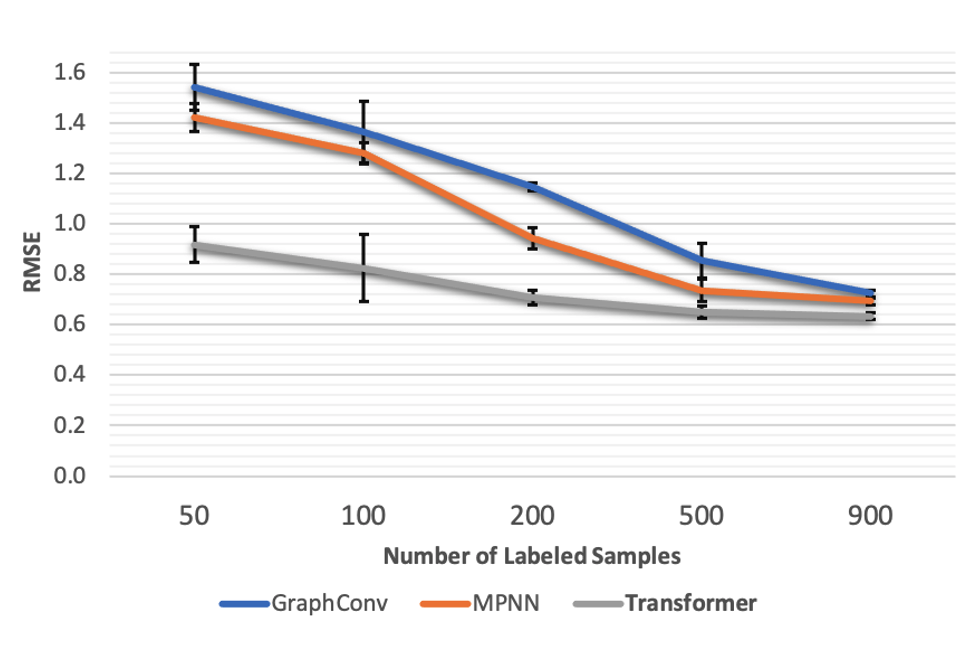}
    \caption{ESOL}
    \label{esol_limit}
  \end{subfigure}%
  ~
  \begin{subfigure}[t]{0.5\columnwidth}
    \centering \includegraphics[scale=0.4]{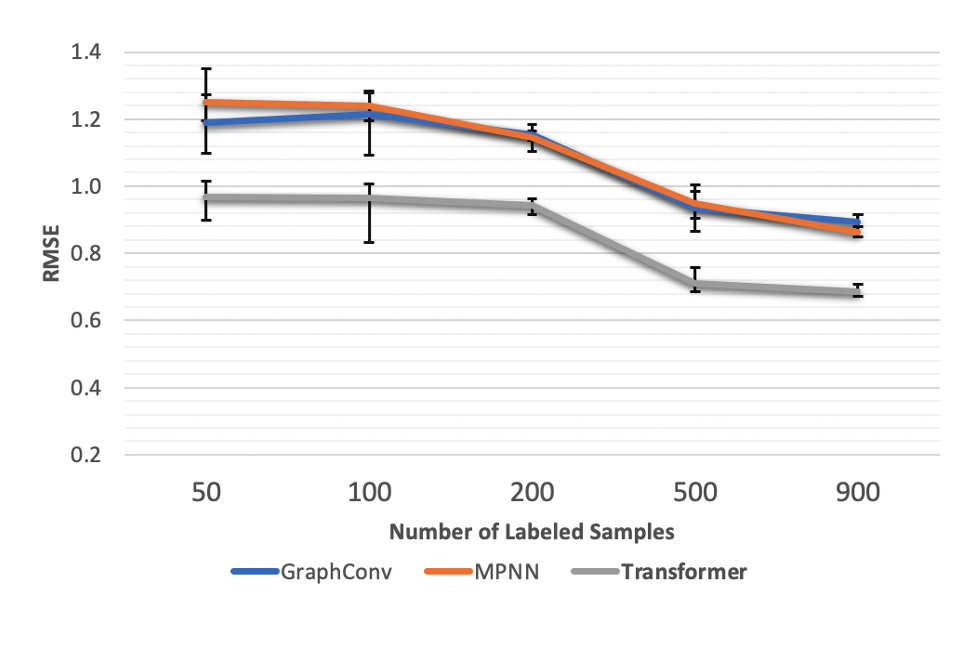}
    \caption{Lipophilicity}
    \label{lipo_limit}
  \end{subfigure}%
  \caption{Label ablation experiment varying number of sample labels seen during
    training} \footnotesize{MPNN and GraphConv implementations are from Wu et
      al.\cite{moleculenet}.}
  \label{ablation}
\end{figure*}
In both of the property prediction tasks, we compare against six different
models from the DeepChem\cite{moleculenet} model library and two models from the
work Tang et al.\cite{Tang} that is closest to our approach, as shown in
Fig. \ref{test_rmse}. We see than graph based models generally outperform the
baseline models. For solubility prediction on ESOL data the best among the
DeepChem\cite{moleculenet} models are the MPNN ($0.61 \pm 0.02$) and Weave
($0.61 \pm 0.07$); our Transformer model ($0.54 \pm 0.017$) outperforms
these models as well as the self-attention MPNN model SAMPN\cite{Tang}
($0.71 \pm 0.01$). Similar behavior is seen for Lipophilicity prediction (fig.\ref{lipo_rmse}) the
attention mechanism in SAMPN does give better performance on RMSE metric (SAMPN:
$0.68 \pm 0.01$) compared to the DeepChem models; however our proposed
Transformer model gives the best performance RMSE
(Transformer $0.60 \pm 0.03$).
For the label ablation study, our Transformer-based model was compared to MPNN and GC. From fig\ref{esol_limit} and
\ref{lipo_limit}, it can be seen that the Transformer model has a much lower RMSE than the other two (Transformer:
$1.01$ on ESOL and $0.90$ on Lipophilicity) in the scenario with the least available target data. The semi-supervised
learning ability helps generalize the molecule encoding space well with fewer labeled targets in this case.
\begin{table*}[h!]
  \centering
  \begin{tabular}{cccc|cc}
      \hline
      cLogP and MR & Neighbor-Atom Prediction & Feature Prediction & Transformer & Test RMSE \\
      Targets & Loss & Loss & STDEV-Update \\
      \hline
      \checkmark & \checkmark & \checkmark & \checkmark & $0.54 \pm 0.01$ \\ 
      \hline
      X & \checkmark & \checkmark & \checkmark & $0.65 \pm 0.10$ \\
      \hline
      \checkmark & X  & \checkmark & \checkmark & $0.58 \pm 0.04$ \\
      \hline
      \checkmark & \checkmark & X & \checkmark & $0.64 \pm 0.08$ \\
      \hline
      \checkmark & \checkmark & \checkmark & X & $0.70 \pm 0.04$ \\      
  \end{tabular}
  \caption{Model performance on ESOL (full data) with individual components removed.} 
  \label{ablation_table}
  \vspace{-1em}
\end{table*}

\autoref{ablation_table} shows performance variation when critical components of the proposed Transformer model are
ablated. We see that the RMSE drops to $0.65 \pm 0.10$ when the RDKit computed targets are ablated. Also, ablating the
standard deviation term in the update drops the RMSE to $0.70 \pm 0.04$.

\section{Conclusion}
\label{conclude}
We have proposed a Transformer-based model for predicting molecular properties. Our approach has better performance on
Lipophilicity and Solubility than state of the art graph-based models and a recent attention-based Message-passing
Network.  Through target label ablation experiments this paper highlights that applying semi-supervised training to
Transformer models can result in good generalization from as few as 50 labeled molecules, unlike other purely supervised
techniques. This work also brings to light the importance of models that work well in settings with minimal target data
but virtually unlimited access to potential molecules, a scenario which is common in the early stages of the drug
discovery process. Transformers often require substantially more data and time for pre-training than was needed here,
mostly because we are only using the Transformer to predict molecule level properties. We suspect that accurately
capturing lower level behavior will require substantial additional data using pre-training such as training in an
auto-encoder setting, BERT-like training to reproduce missing atom vectors, and/or training using a large number of
similar tasks and targets, such as training to reproduce the broad set of assays in e.g. BindingDB \cite{bdb}.

\section{Acknowledgments}
We thank Murat Sorkun of DIFFER/AMD, Netherlands, for valuable inputs on the experimental results \cite{sorkun}.
This material is based upon work supported by the Defense Advanced Research Projects Agency (DARPA) under Contract
No. HR001119C0108. The views, opinions, and/or findings expressed are those of the author(s) and should not be
interpreted as representing the official views or policies of the Department of Defense or the U.S. Government (Approved
for public release, distribution unlimited).

{\small \bibliographystyle{ieee_fullname} \bibliography{neurips_2020} }

\begin{thebibliography}{10}\itemsep=-1pt

\bibitem{transformerXL}
Zihang Dai, Zhilin Yang, Yiming Yang, Jaime Carbonell, Quoc~V Le, and Ruslan
  Salakhutdinov.
\newblock Transformer-xl: Attentive language models beyond a fixed-length
  context.
\newblock {\em arXiv preprint arXiv:1901.02860}, 2019.

\bibitem{esol}
John~S. Delaney.
\newblock Esol:  estimating aqueous solubility directly from molecular
  structure.
\newblock {\em Journal of Chemical Information and Computer Sciences},
  44(3):1000--1005, 2004.

\bibitem{bert}
Jacob Devlin, Ming-Wei Chang, Kenton Lee, and Kristina Toutanova.
\newblock Bert: Pre-training of deep bidirectional transformers for language
  understanding.
\newblock {\em arXiv preprint arXiv:1810.04805}, 2018.

\bibitem{MPNN}
Justin Gilmer, Samuel~S Schoenholz, Patrick~F Riley, Oriol Vinyals, and
  George~E Dahl.
\newblock Neural message passing for quantum chemistry.
\newblock {\em arXiv preprint arXiv:1704.01212}, 2017.

\bibitem{weave}
Steven Kearnes, Kevin McCloskey, Marc Berndl, Vijay Pande, and Patrick Riley.
\newblock Molecular graph convolutions: moving beyond fingerprints.
\newblock {\em Journal of computer-aided molecular design}, 30(8):595--608,
  2016.

\bibitem{bdb}
Tiqing Liu, Yuhmei Lin, Xin Wen, Robert~N Jorissen, and Michael~K Gilson.
\newblock Bindingdb: a web-accessible database of experimentally determined
  protein--ligand binding affinities.
\newblock {\em Nucleic acids research}, 35(suppl\_1):D198--D201, 2007.

\bibitem{ecfp}
David Rogers and Mathew Hahn.
\newblock Extended-connectivity fingerprints.
\newblock {\em Journal of Chemical Information and Modeling}, 50(5):742--754,
  2010.

\bibitem{shaw18}
Peter Shaw, Jakob Uszkoreit, and Ashish Vaswani.
\newblock Self-attention with relative position representations.
\newblock In {\em Proceedings of the 2018 Conference of the North {A}merican
  Chapter of the Association for Computational Linguistics: Human Language
  Technologies, Volume 2 (Short Papers)}, pages 464--468, New Orleans,
  Louisiana, June 2018. Association for Computational Linguistics.

\bibitem{shin}
Bonggun Shin, Sungsoo Park, Keunsoo Kang, and Joyce~C. Ho.
\newblock Self-attention based molecule representation for predicting
  drug-target interaction.
\newblock In Finale Doshi{-}Velez, Jim Fackler, Ken Jung, David~C. Kale, Rajesh
  Ranganath, Byron~C. Wallace, and Jenna Wiens, editors, {\em Proceedings of
  the Machine Learning for Healthcare Conference, {MLHC} 2019, 9-10 August
  2019, Ann Arbor, Michigan, {USA}}, volume 106 of {\em Proceedings of Machine
  Learning Research}, pages 230--248. {PMLR}, 2019.

\bibitem{sorkun}
Murat Sorkun, J.~M.~Vianney~A. Koelman, and S{\"u}leyman Er.
\newblock Pushing the limits of solubility prediction via quality-oriented data
  selection.
\newblock {\em Research Square}, 2020.

\bibitem{Tang}
Bowen Tang, Skyler Kramer, Meijuan Fang, Zhen Wu, and Dong Xu.
\newblock A self-attention based message passing neural network for predicting
  molecular lipophilicity and aqueous solubility.
\newblock {\em Journal of Cheminformatics}, 12, 12 2020.

\bibitem{transformer}
Ashish Vaswani, Noam Shazeer, Niki Parmar, Jakob Uszkoreit, Llion Jones,
  Aidan~N Gomez, {\L}ukasz Kaiser, and Illia Polosukhin.
\newblock Attention is all you need.
\newblock In {\em Advances in neural information processing systems}, pages
  5998--6008, 2017.

\bibitem{moleculenet}
Zhenqin Wu, Bharath Ramsundar, Evan N. Feinberg, Joseph Gomes, Caleb Geniesse,
  Aneesh~S. Pappu, Karl Leswing, and Vijay Pande.
\newblock Moleculenet: a benchmark for molecular machine learning.
\newblock {\em Chem. Sci.}, 9:513--530, 2018.

\bibitem{mit}
Kevin Yang, Kyle Swanson, Wengong Jin, Connor Coley, Philipp Eiden, Hua Gao,
  Angel Guzman-Perez, Timothy Hopper, Brian Kelley, Miriam Mathea, Andrew
  Palmer, Volker Settels, Tommi Jaakkola, Klavs Jensen, and Regina Barzilay.
\newblock Analyzing learned molecular representations for property prediction.
\newblock {\em Journal of Chemical Information and Modeling}, 59(8):3370--3388,
  2019.

\bibitem{smilessyn}
Shuangjia Zheng, Xin Yan, Yuedong Yang, and Jun Xu.
\newblock Identifying structure–property relationships through smiles syntax
  analysis with self-attention mechanism.
\newblock {\em Journal of Chemical Information and Modeling}, 59(2):914--923,
  2019.

\end{thebibliography}

\end{document}